\documentclass[aps,twocolumn,superscriptaddress,showpacs]{revtex4}

\usepackage{graphics}
\usepackage{graphicx}
\usepackage{bm}

\begin{document}

\title{Tunable electronic transport and unidirectional quantum wires in
graphene subjected to electric and magnetic fields}
\author{Yury P. Bliokh}
\affiliation{Advanced Science Institute, The Institute of Physical and Chemical Research
(RIKEN), Wako-shi, Saitama 351-0198, Japan}
\affiliation{Department of Physics , Technion-Israel Institute of Technology, Haifa 32000,
Israel}
\author{Valentin Freilikher}
\affiliation{Advanced Science Institute, The Institute of Physical and Chemical Research
(RIKEN), Wako-shi, Saitama 351-0198, Japan}
\affiliation{Jack and Pearl Resnick Institute of Advanced Technology, Department of
Physics, Bar-Ilan University, Ramat-Gan 52900, Israel}
\author{Franco Nori}
\affiliation{Advanced Science Institute, The Institute of Physical and Chemical Research (RIKEN), Wako-shi, Saitama 351-0198, Japan}
\affiliation{Department of Physics, University of Michigan, Ann Arbor, Michigan 48109-1040, USA. }

\begin{abstract}
Magnetic barriers in graphene are not easily tunable. Here we show that the
application of both electric and magnetic fields provides tunable and far
more controllable electronic states in graphene. In particular, a
one-dimensional channel (quantum wire) can be created, which supports
localized electron-hole states with parameters tunable by the electric field.
Such quantum wire offers peculiar conducting properties, like unidirectional
conductivity and robustness to disorder. Two separate quantum wires comprise
a waveguide with two types of eigenmodes: one type is similar to traditional
waveguides, the other type is formed by coupled surface waves propagating
along the boundaries of the waveguide.
\end{abstract}

\maketitle

\section{Introduction}

The realization of stable single-layer carbon crystals (graphene) triggered an explosion of interest in this material because of its unique electronic properties (for reviews see, e.g., \cite{Castro1, Geim1, Novoselov2}), making it a promising candidate for designing one-chip nanoelectronic devices (e.g., \cite{Geim1, Katsnelson, Trauzettel,Berger, Son}). However, Klein tunneling \cite{Klein} hinders the application of traditional methods of current control (e.g., on-off switching, changing the current direction, etc.) by tuning the voltage between various elements of a device  \cite{Katsnelson2}. This effect also complicates the creation of localized
electron-hole states in graphene.

As it was shown recently \cite{DeMartino1,DeMartino2}, this difficulty in creating
localized states can be overcome by using an inhomogeneous magnetic field.
An alternative way (which is mathematically analogous to the previous one),
involving graphene sheet deformations, was proposed in \cite{Pereira, Guinea,Guinea-2}. The regions with either inhomogeneous magnetic field or strains can act as
non-transparent barriers and angle-resolved charge carriers filters \cite{Masir2} and form long-lived \cite{Masir1} or stationary localized states 
\cite{DeMartino1, DeMartino2, Kormanyos, Park}. These 1-D states appear as
``surface'' waves propagating along the
inhomogeneity of the magnetic field \cite{Muller, Orozlany, Ghosh1,
Peeters1, Peeters2, Peeters3}, or along the strains \cite{Pereira}.

The charge-confining and guiding capabilities of magnetic barriers
\cite{DeMartino1, DeMartino2} and graphene strains \cite{Pereira} 
open up certain possibilities for potential future applications. However
when it comes to designing fast-tunable electronic devices (switches,
filters, etc.) a difficulty emerges. The problem is that most of the
existing magnetic barrier technologies usually imply the deposition (on top
or beneath the graphene sheet) of a \textit{fixed} pattern of magnetic
material which reproduces the desired magnetic field distribution in the
sample. Any change of parameters means in fact building a new setup and
creates formidable (if surmountable) obstacles for harnessing magnetic
barriers as elements of fast-acting electronic devices. In other words,
magnetic barriers are not easily tunable.

Here we suggest an efficient way around this problem by simultaneously
employing both inhomogeneous magnetic and electric fields: i.e., \textit{combined electric and magnetic barriers}. The proper combination of these two allows a better control of the transport properties of graphene
by tuning the electric potential, with the parameters of the magnetic field
remaining intact. Depending on the voltage, this barrier can be either
semitransparent or opaque.

The combined electrostatic and magnetic barrier possesses a unique feature
that makes it different from other types of barriers. Graphene with mutually
perpendicular electric and magnetic fields supports states which are
localized near the barrier. These current-carrying states (surface waves)
correspond to quasiparticles moving along the barrier just in one direction.
The direction and the value of the quasiparticle velocity is easily
controlled by the electrostatic potential. These states correspond to the
classical drift of charged particles in crossed electric and magnetic fields
and exist if and only if the drift velocity is smaller than the Fermi
velocity. The absence of counter-propagating states prevents the
backscattering induced by either irregularities in graphene \cite{Titov,
BliokhPRB, Rozhkov} or by the fluctuations of the magnetic field. For more
about the effects of disorder on the electronic properties of graphene see
Chapter IV in [1] and references therein.

While one barrier forms a wire, two such barriers make up a waveguide. This
waveguide has a set of ``ordinary'' waves
and another set of ``extraordinary'' waves.
The ordinary waves are characterized by the quantization of their transverse
wave numbers, while the extraordinary waves are formed by two coupled
surface waves propagating along the waveguide walls (barriers). Depending on
the barrier parameters, the extraordinary modes can be either bidirectional
or unidirectional. There is an energy gap where only extraordinary modes
exist. Decreasing the spacing between the barriers broadens this gap. The
extraordinary modes are also stable against backscattering.

The unique, easily controllable, and tunable features of the combined
barriers and waveguides are promising toward creating new graphene-based
electronic devices.

This paper is organized as follows. In Sec.~II, the models for the
electromagnetic barriers and auxiliary constructions are described. In
Sec.~III, the properties of a single electromagnetic barrier are studied.
The spectrum and eigenfunction of the waveguide formed by two barriers are
presented in Sec.~IV. Concluding remarks are in Sec.~V.

\section{Basic formalism}

\subsection{Step-like barrier.}

The low-energy excitations in single-layer graphene in the presence of a
perpendicular to the layer magnetic field $H_{z}(x)=dA_{y}/dx$ and an
in-plane electric field $E_{x}(x)=-dV(x)/dx$ (which are constant along the $%
y $-direction) are described by the Dirac equations: 
\begin{eqnarray}
\left[ -i{\frac{\partial }{\partial x}}+{\frac{\partial }{\partial y}}+{%
\frac{ie}{c\hbar }}A(x)\right] \psi _{A} &=&{\frac{1}{v_{F}\hbar }}[\mathcal{%
E}-eV(x)]\psi _{B},  \nonumber  \label{eq1} \\
\left[ -i{\frac{\partial }{\partial x}}-{\frac{\partial }{\partial y}}-{%
\frac{ie}{c\hbar }}A(x)\right] \psi _{B} &=&{\frac{1}{v_{F}\hbar }}[\mathcal{%
E}-eV(x)]\psi _{A}.
\end{eqnarray}%
where $A(x)$ is the $y$-component of the vector potential, $V(x)$ is the
scalar potential, $\mathcal{E}$ is the energy of the quasiparticle, and $%
\psi $ is the two-component spinor $\psi =(\psi _{A},\,\psi _{B})^{T}$.

\begin{figure}[tbh]
\centering \scalebox{1.0}{\includegraphics{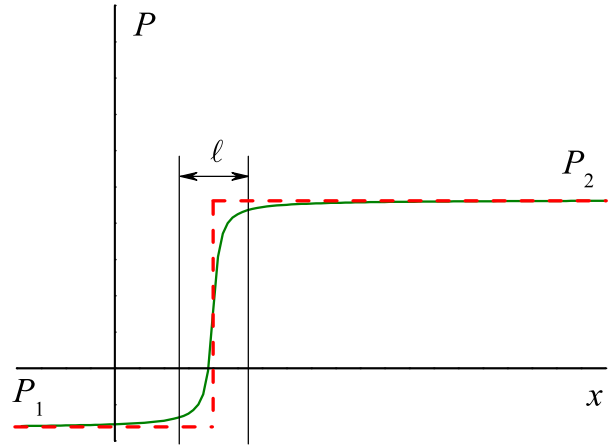}}
\caption{(color online) Schematic diagram of the inhomogeneity region
between two otherwise-homogeneous graphene sheets. The function $P(x)$ can
be either the electric scalar $V(x)$ or magnetic $A(x)$ potential (hence the
choice of $P$). Here, $P(x\rightarrow -\infty)\rightarrow P_1$, and $%
P(x\rightarrow \infty)\rightarrow P_2$. The dashed line is the step-like
barrier approximation. Note that $\bm{A}\equiv A_y\hat{\bm{y}}\equiv A(x)%
\hat{\bm{y}}$.}
\label{Fig1}
\end{figure}

Let us consider two homogeneous graphene domains subjected to different
constant{\large \ }scalar ($V_{1}$ and $V_{2}$) and vector ($A_{y1}\equiv
A_{1}$ and $A_{y2}\equiv A_{2}$) potentials, and assume that the domains are
connected by an inhomogeneous transition region where the potentials vary
smoothly between the constant values (see Fig.~\ref{Fig1}). When the width $%
\ell $ of this region is large compared to the graphene lattice spacing $a$,
and small compared to the Fermi wavelength $\lambda _{F}$ ($a\ll \ell \ll
\lambda _{F}$), the potentials can be replaced by step-like functions, as
shown by the red dashed line in Fig.~\ref{Fig1}. This widely used
approximation simplifies the problem considerably.

Due to translational invariance along the $y$-direction, the solutions of
Eqs.~(\ref{eq1}) can be presented in the form $\psi_{A,B}(x,y)=\exp(ik_yy)%
\Psi_{A,B}(x)$. In dimensionless variables $\xi=x/L$, 
\[ \kappa_\perp=k_y L,\;\; \varepsilon=\mathcal{E}L/\hbar v_F,\]
\[u=eVL/\hbar v_F,\;\; \mathcal{A}=(eL/c\hbar)A,\]
where $L$ is a characteristic spatial scale (magnetic length $\ell_H=\sqrt{%
c\hbar/eH}$, for instance), Eqs.~(\ref{eq1}) have the form: 
\begin{eqnarray}  \label{eq2}
\left({\frac{d}{d\xi}}-\kappa_\perp-\mathcal{A}\right)\psi_A=i(%
\varepsilon-u)\psi_B,  \nonumber \\
\left({\frac{d}{d\xi}}+\kappa_\perp+\mathcal{A}\right)\psi_B=i(%
\varepsilon-u)\psi_A.
\end{eqnarray}

\subsection{Homogeneous graphene.}

\begin{figure}[tbh]
\centering \scalebox{0.4}{\includegraphics{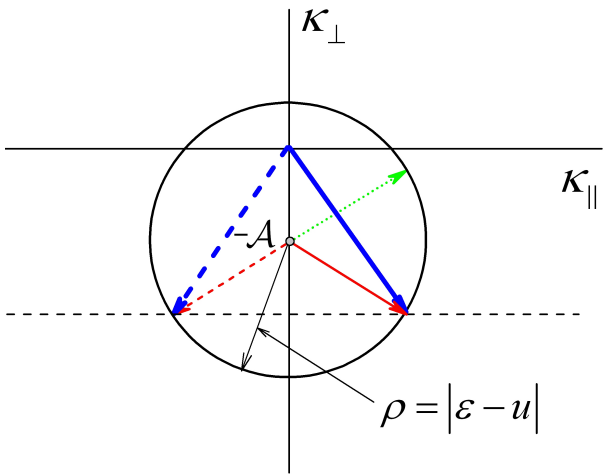}}
\caption{(color online) The wave vectors $\bm{\kappa}$ (blue thick arrows) of
propagating waves in homogeneous graphene lie on a circle, described by the
dispersion equation (\ref{eq3}). The normalized coefficients $\rho 
C^{(+)}$ (red solid arrow) and $\rho C^{(-)}$ (red dashed arrow) are
shown when $\mathrm{sgn}(\varepsilon-u)=+1$. The current with
positive $J_x$ component is directed either along the solid red arrow when $\mathrm{sgn}(\varepsilon-u)=+1$, or along the green dotted arrow when $\mathrm{sgn}(\varepsilon-u)=-1$.}
\label{Fig2}
\end{figure}

In homogeneous graphene ($\mathcal{A}=\mathrm{constant}$, $u=\mathrm{constant}$), $\Psi _{A,B}(\xi)\sim \exp (i\kappa _{\parallel }\xi )$, and the wave
vector components $\kappa _{\parallel }$ and $\kappa _{\perp }$ are related
by the dispersion relation: 
\begin{equation}
\kappa _{\parallel }^{2}+(\kappa _{\perp }+\mathcal{A})^{2}=(\varepsilon
-u)^{2}.  \label{eq3}
\end{equation}%
This equation is valid for both propagating ($\mathrm{Im}\kappa _{\parallel
}=0$) and non-propagating (evanescent) waves ($\mathrm{Re}\kappa _{\parallel
}=0$). In the $(\kappa _{\parallel },\kappa _{\perp })$-plane, the wave
vectors $\bm{\kappa}$ of the propagating waves lie on a circle centered at
the point $\kappa _{\perp }=-\mathcal{A}$, with radius $\rho
=|\varepsilon -u|$ (Fig.~\ref{Fig2}). For given $\kappa _{\perp }$ and $\varepsilon $, there are two solutions of the dispersion equation (\ref{eq3}) with positive (blue solid arrow in Fig.~\ref{Fig2}) and negative (blue
dashed arrow in Fig.~\ref{Fig2}) values of $\kappa _{\parallel }$.

\subsection{Probability current.}

The solutions of Eq.~(\ref{eq2}) have the form: 
\begin{equation}
\Psi _{A,B}=\psi _{A,B}^{(+)}e^{i\kappa _{\parallel }\xi }+\psi
_{A,B}^{(-)}e^{-i\kappa _{\parallel }\xi }.  \label{eq4}
\end{equation}%
The amplitudes $\psi _{A,B}^{(\pm )}$ are connected by the relation: 
\begin{equation}
\psi _{B}^{(\pm )}=C^{(\pm )}\psi _{A}^{(\pm )},  \label{eq5}
\end{equation}%
where 
\begin{equation}
C^{(\pm )}={\frac{i(\kappa _{\perp }+\mathcal{A})\pm \kappa _{\parallel }}{%
\varepsilon -u}},  \label{eq6}
\end{equation}%
and 
\begin{equation}
C^{(-)}=-1/C^{(+)}.  \label{eq7}
\end{equation}%
Eq.~(\ref{eq5}) allows the consideration of the spinor component $\Psi _{A}$
only (thereafter $\Psi $). When the wave vector $\bm{\kappa}$ lies on the
circle (propagating waves), the following representation of the coefficients 
$C^{(\pm )}$ is valid: 
\begin{equation}
C^{(+)}\equiv C=\mathrm{sgn}(\varepsilon -u)e^{i\varphi },\hspace{3mm}%
C^{(-)}=-C^{\ast },  \label{eq8}
\end{equation}%
where 
\begin{equation}
\sin \varphi ={\frac{\kappa _{\perp }+\mathcal{A}}{|\varepsilon -u|}},%
\hspace{5mm}\cos \varphi ={\frac{\kappa _{\parallel }}{|\varepsilon -u|}}.
\label{eq9}
\end{equation}%
The angle $\varphi $ is connected with the direction of the probability
current density, $\bm{J}$ \cite{DeMartino1, DeMartino2}. Indeed, the probability current
density $\bm{J}=\langle \psi |\bm{\sigma}|\psi \rangle $ can be presented in
the form: 
\begin{equation}
J_{x}+iJ_{y}=2\psi _{A}^{\ast }\psi _{B}.  \label{eq10}
\end{equation}%
Using Eq.~(\ref{eq10}), the current densities $\bm{J}^{(\pm )}$ that
correspond to pure $(+)$ and $(-)$ states, can be written as: 
\begin{eqnarray}
\bm{J}^{(+)} &=&2\,\mathrm{sgn}(\varepsilon
-u)e^{i\varphi }\left\vert \psi ^{(+)}\right\vert ^{2},  \nonumber
\label{eq11} \\
\bm{J}^{(-)} &=&-2\,\mathrm{sgn}(\varepsilon -u)e^{-i\varphi }\left\vert
\psi ^{(-)}\right\vert ^{2}.
\end{eqnarray}%
It follows from Eq.~(\ref{eq11}) that the current with positive component $%
J_{x}>0$ is described by $\psi ^{(+)}$ when $\mathrm{sgn}(\varepsilon -u)=+1$
(red arrow in Fig.~\ref{Fig1}), and by $\psi ^{(-)}$ when $\mathrm{sgn}%
(\varepsilon -u)=-1$ (green arrow in Fig.~\ref{Fig1}).

\section{Combined electric and magnetic barrier}

\subsection{Semi-transparent barrier.}

The continuity condition for the spinor components $\Psi_{A,B}$ on the
interface between two homogeneous domains (domain 1 and domain 2) can be
written in the form: 
\begin{equation}  \label{eq12}
(\psi^{(+)}_2, \psi^{(-)}_2)^T=\hat{M}(\psi^{(+)}_1, \psi^{(-)}_1)^T,
\end{equation}
where $\hat{M}$ is the transfer matrix: 
\begin{equation}  \label{eq13}
\hat{M}={\frac{1}{C_{2}+C_{2}^{-1}}}\left\Vert 
\begin{array}{cc}
C_{2}^{-1}+C_1 & C_{2}^{-1}-C_1^{-1} \\ 
C_{2}-C_1 & C_{2}+C_1^{-1}%
\end{array}
\right\Vert.
\end{equation}
The coefficients $C_{1,2}$ are defined by Eqs.~(\ref{eq6})--(\ref{eq8}),
with potentials $u_{1,2}$ and $\mathcal{A}_{1,2}$, correspondingly.

\begin{figure}[tbh]
\centering \scalebox{0.4}{\includegraphics{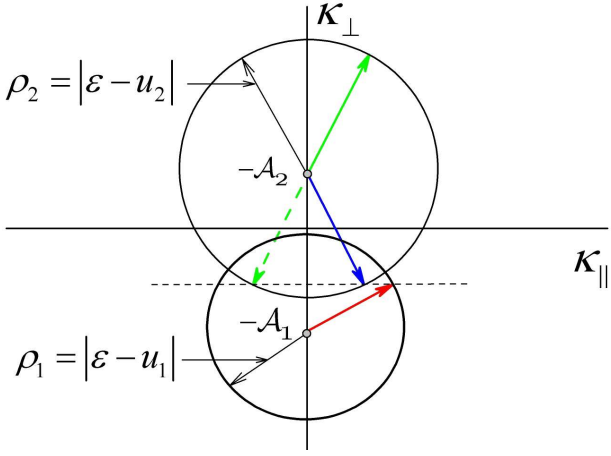}}
\caption{(color online) Here $\kappa_\perp$ and $\kappa_\parallel$ are the transverse and longitudinal components of the wave vector $\bm{\kappa}$. The circles represent solutions of the dispersion relations $\varepsilon(\bm{\kappa})$ in the domains 1 and 2, and correspond to propagating waves. The top circle has a radius $\rho_2=|\varepsilon-u_2|$, while the bottom circle has a radius $\rho_2=|\varepsilon-u_2|$. 
The incident current is directed along the bottom-right red arrow. The
refracted current is directed along the blue arrow (the only arrow pointing
South-East, towards 5 o'clock), when $\mathrm{sgn}(\varepsilon-u_2)=+1$, and along the green solid arrow (pointing North-East, 1 o'clock)
when $\mathrm{sgn}(\varepsilon-u_2)=-1$. The latter one is opposite
to the dashed (South-West, 7 o'clock) green one, which intersects the black
horizontal dashed line.}
\label{Fig3}
\end{figure}

To describe the transport properties at the barrier, we now
introduce a graphic representation, shown in Fig.~\ref{Fig3}, which provides
a better understanding of the dispersion relations. In Fig.~\ref{Fig3},
points on the circles represent solutions of the dispersion equations in
domains 1 and 2, and correspond to propagating waves (waves with real $\kappa _{\parallel }$) in these domains.

Let us now consider the ``refraction law''
which relates the directions of the incident and refracted waves [states
with the same sign of the current component $J_{x}$\ in both domains (for
definiteness, $J_{x}>0$)]. This ``refraction law'' can be easily derived with the help of Fig.~\ref{Fig3}
as follows. Due to translation invariance along the $y$-axis, the wave
vector component $\kappa _{\perp }$ has the same value in both media. The
wave vectors $\bm{\kappa}$ of the propagating waves take on values $%
\bm{\kappa}_{1}$ and $\bm{\kappa}_{2}$ in the first and second domains. The
wave vectors $\bm{\kappa}_{1}$ and $\bm{\kappa}_{2}$ lie at the intersection
of the corresponding circles and the line $\kappa _{\perp }=\mathrm{constant}
$ (thin horizontal dashed black line in Fig.~\ref{Fig3}). There are two such
intersections for each of the two circles. The physically meaningfull
intersections (i.e., the intersections that present the solutions with
positive current density component $J_{x}>0$) are defined by the sign of $%
(\varepsilon -u_{1,2})$. The currents $\bm{J}$ are directed either along the
radius vector (i.e., the vectors from the centers of both circles to the
intersections, in Fig.~\ref{Fig3}) of the intersection or in the opposite
direction, depending on the sign of: $(\varepsilon -u_{1,2})$. For example,
when $\mathrm{sgn}(\varepsilon -u_{1})=+1$, the incident current $\bm{J}$ is
directed along the red arrow in Fig.~\ref{Fig3}. The refracted current is
directed along the blue arrow when $\mathrm{sgn}(\varepsilon -u_{2})=+1$, or
along the green arrow when $\mathrm{sgn}(\varepsilon -u_{2})=-1$.

The following relation defines the connection between the incident, $\theta
_{1}$, and refracted, $\theta _{2}$, angles: 
\begin{equation}
\kappa _{\perp }=-\mathcal{A}_{1}-(\varepsilon-u_{1})\sin \theta _{1}=-%
\mathcal{A}_{2}-(\varepsilon -u_{2})\sin \theta _{2}.  \label{eq13a}
\end{equation}%
When, forinstance, $|\varepsilon -u_{1}|=|\varepsilon -u_{2}|\equiv
|\varepsilon -u|$, the refraction law reads: 
\begin{equation}
\mathrm{sgn}(\varepsilon -u_{1})\sin \theta _{1}+\mathrm{sgn}(\varepsilon
-u_{2})\sin \theta _{2}={\frac{\mathcal{A}_{2}-\mathcal{A}_{1}}{|\varepsilon
-u|}}.  \label{eq13b}
\end{equation}%
The refraction law presented in \cite{DeMartino1, DeMartino2} is a particular case $%
(u_{1}=u_{2}=0)$ of this relation.

It is possible to show that the transmission coefficient $T=J_{x\,1}/J_{x\,2} $ is defined by the following expression: 
\begin{equation}
T\equiv T\left( w_{2},w_{1}\right) ={\frac{\textstyle2\sqrt{1-w_{2}^{2}}\sqrt{1-w_{1}^{2}}}{\textstyle1+\sqrt{1-w_{2}^{2}}\sqrt{1-w_{1}^{2}}-w_{2}w_{1}}},  \label{eq14}
\end{equation}%
where 
\[w_{2,1}={\frac{\textstyle\kappa _{\perp }+\mathcal{A}_{2,1}}{\textstyle\varepsilon -u_{2,1}}}.\]
It follows from Eq.~(\ref{eq14}) that 
\begin{equation}
\left. T\right\vert _{w_{2}=w_{1}}=1,  \label{eq15}
\end{equation}%
i.e., there is an angle of incidence, non-normal in the general case, for
which the interface is totally transparent. In other words, the
barrier changes the direction of incidence, at which Klein tunneling occurs 
\cite{Feng}.

\subsection{Non-transparent barrier: bound states.}

When the difference between the vector potentials is large enough, 
\begin{equation}
|\mathcal{A}_{1}-\mathcal{A}_{2}|>|\varepsilon -u_{1}|+|\varepsilon -u_{2}|,
\label{eq16}
\end{equation}%
the circles in Fig.~\ref{Fig3} do not cross, and no propagating wave can
penetrate through the interface. Therefore, the transmission coefficient is
equal to zero, $T=0$, and the barrier is nontransparent (reflecting wall).
However, it can support a wave, which propagates \textit{along} the line
separating two domains with the amplitude exponentially decaying in the
transverse directions. This mode is a 1D analog of two-dimensional surface
waves. In order to simplify terminology, in what follows, we call it a
surface wave.

Let us now determine the properties and existence conditions of these
surface waves. When the wave vector components $\kappa _{\parallel \,j}$ ($%
j=1,2$) in both media are imaginary, $\kappa _{\parallel \,j}=i|\kappa
_{\parallel \,j}|$ the wave functions $\psi _{1,2}$ have the form: 
\begin{equation}
\psi _{j}=\psi _{j}^{(+)}e^{-|\kappa _{\parallel \,j}|\xi }+\psi
_{j}^{(-)}e^{+|\kappa _{\parallel \,j}|\xi },  \label{eq17}
\end{equation}%
where $\kappa _{\parallel \,j}=\sqrt{(\varepsilon -u_{j})^{2}-\kappa _{\perp
}^{2}}=i\sqrt{\kappa _{\perp }^{2}-(\varepsilon -u_{j})^{2}}$. The wave is
now localized near the interface{\Huge \ } $\xi =0$ when $\psi
_{1}^{(-)}=\psi _{2}^{(+)}=0$, i.e. $\psi _{2}^{(+)}=M_{11}\psi _{1}^{(+)}=0$%
. Therefore, the condition 
\begin{equation}
M_{11}={\frac{C_{1}C_{2}+1}{C_{2}^{2}+1}}=0  \label{eq18}
\end{equation}%
is the dispersion relation of the surface waves. A solution $\varepsilon
(\kappa _{\perp })$ of this equation exists if and only if 
\begin{equation}
\left\vert {\frac{u_{1}-u_{2}}{\mathcal{A}_{1}-\mathcal{A}_{2}}}\right\vert
<1  \label{eq19}
\end{equation}%
and has the form: 
\begin{equation}
\varepsilon (\kappa _{\perp })={\frac{u_{1}\mathcal{A}_{2}-u_{2}\mathcal{A}%
_{1}}{\mathcal{A}_{2}-\mathcal{A}_{1}}}+\kappa _{\perp }{\frac{u_{1}-u_{2}}{%
\mathcal{A}_{2}-\mathcal{A}_{1}}}.  \label{eq20}
\end{equation}%
Equation~(\ref{eq20}) describes a surface wave propagating along the line $%
\xi =0$, with the group velocity $\nu _{g}\sim d\varepsilon /d\kappa _{\perp
}=(u_{1}-u_{2})/(\mathcal{A}_{2}-\mathcal{A}_{1})$.

Note that the inequality (\ref{eq19}) in the dimensional variables takes the
form: 
\begin{equation}
{\frac{c}{v_{F}}}\left\vert {\frac{\Delta V}{\Delta A}}\right\vert <1,
\label{eq19a}
\end{equation}%
where $\Delta V$ and $\Delta A$ are, respectively, the differences between
the scalar and vector potentials in the two graphene domains. From Fig.~\ref%
{Fig1} one can see that the electric field in the inhomogeneous region is $%
E_{x}\simeq -\Delta V/\ell $, and the magnetic field is $H_{z}\simeq \Delta
A/\ell $. Therefore, the inequality (\ref{eq19a}) can be written as: 
\begin{equation}
v_{d}\equiv c\left\vert {\frac{E_{x}}{H_{z}}}\right\vert <v_{F}.
\label{eq19b}
\end{equation}%
Thus, the dimensional group velocity $v_{g}=\hbar^{-1}d\mathcal{E}/dk_{y}$ of the
surface wave, coincides with the \textit{drift velocity} $v_{d}=cE/H$ of a
charged particle in crossed electric and magnetic fields.

If, for example, the characteristic width, $\ell$, of the
barrier (see Fig.~\ref{Fig1}) is $\ell =10$nm, and the magnetic field $H_{z}=1$~T, the condition for the surface wave to exist,$v_{d}<v_{F}$ [Eq.~(\ref{eq19b})], is satisfied when the potential
difference across the barrier (between two graphene domains in Fig.~\ref{Fig1})
is $\Delta V=10$mV or less. These numbers are quite feasible.
Magnetic barriers with amplitudes of up 1~T have been created experimentally
by depositing ferromagnetic films on top of a graphene sheet \cite{Kubrak,
Crerchez}. Patterened stripes down to 10nm can be realized using
nanolitography (see \cite{Bader} for review).

Note that the bound state near the $\delta$-function magnetic barrier [$%
H_z\sim\delta(x)$] described in \cite{Masir2} is the particular case $%
u_1=u_2 $ ($E_x=0$) of this surface wave with zero group velocity.

\subsection{Regions in the $(\kappa_\perp,\varepsilon)$%
-plane where surface waves exist}

\begin{figure}[tbh]
\centering \scalebox{0.45}{\includegraphics{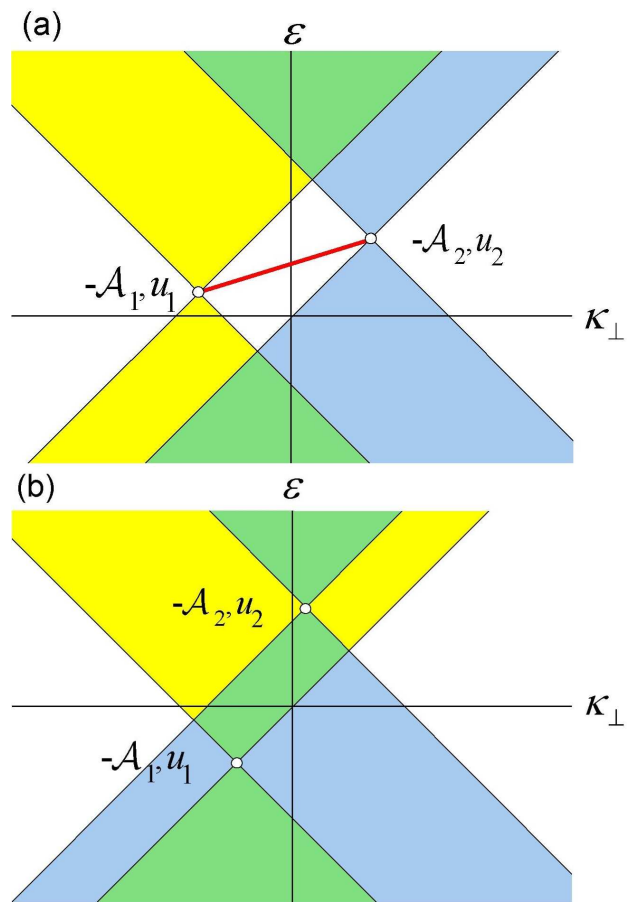}}
\caption{(color online) The yellow and blue sectors correspond to
propagating waves in domains 1 and 2, respectively. The overlapping (green)
regions correspond to waves that can propagate both in domain 1 and domain
2. (a) the inequality Eq.~(\ref{eq19}) is satisfied. The barrier is opaque for all the waves whose energy lies in the energy gap between the green regions. The
dispersion $\varepsilon(\kappa_\perp)$ of the surface wave
is shown by the straight red line joining both Dirac points.  (b) the
inequality Eq.~(\ref{eq19}) is not satisfied and there is no surface wave. The barrier is semitransparent, i.e., regardless of the energy, there are always the angles of incidence when the waves penetrate through the barrier.}
\label{Fig4}
\end{figure}

The regions in the ($\kappa _{\perp },\,\varepsilon $)-plane where the
surface waves exist can be defined using the graphic construction shown in
Fig.~\ref{Fig4}. The Dirac point $(-\mathcal{A}_{1},u_{1})$ in the graphene
domain~1 generates a division of the plane into four sectors. Two of them
(yellow sectors in Fig.~\ref{Fig4}a) correspond to propagating waves with
real $\kappa _{\parallel }$, and the other two sectors correspond to
non-propagating waves with imaginary $\kappa _{\parallel }$ (white regions
in Fig.~\ref{Fig4}a). A similar division of the plane is generated by the
Dirac point $(-\mathcal{A}_{2},u_{2})$ in the domain~2. The blue sectors in
Fig.~\ref{Fig4}a correspond to propagating waves in this domain.

Green regions in Fig.~\ref{Fig4}a show the overlap of the propagating-wave
sectors, i.e., waves whose parameters $\kappa _{\perp }$ and $\varepsilon $
lie in these overlapping green regions can simultaneously propagate in both
domains. For these waves the interface between the graphene domains acts as
a semitransparent barrier.

Yellow (blue) sector corresponds to the waves that can propagate only in
domain~1 (only in domain~2), and the interface between the domains acts as a
non-transparent (reflecting) barrier. Parameters $\kappa _{\perp }$ and $%
\varepsilon $ of the non-propagating surface waves (i.e., waves that are
evanescentin the both domains) belong to the white regions in Fig.~\ref{Fig4}%
a. The red solid line that connects the two Dirac points is the dispersion
curve $\varepsilon (\kappa _{\perp })$ described by Eq.~(\ref{eq20}). It
follows from Eq.~(\ref{eq20}) that the surface waves, if exist, propagate in
only one direction. In other words, \textit{such barrier constitutes a
unidirectional quantum wire.}

When the inequality (\ref{eq19}) is not satisfied, the division of the ($%
\kappa _{\perp },\,\varepsilon $)-plane has different structure, shown in
Fig.~\ref{Fig4}b. Here there is no white region between the Dirac points, i.
e., the surface waves\ are absentg in this case. Another important
difference between these two cases is the following. When the inequality (%
\ref{eq19}) holds, there is an energy gap, in which there are no waves
penetrating through the barrier, and the barrier is opaque for all angles of
incidence (all $\kappa _{\perp }$). Figure~\ref{Fig4}a allows easy
determination of this range of energies. The absence of waves penetrating
through the barrier means that the horizontal line $\varepsilon =const$ does
not cross any green regions. It is readily seen that this condition is
satisfied when the energy lies in the following range: 
\[
{\frac{1}{2}}(u_{1}+u_{2}-|\mathcal{A}_{1}-\mathcal{A}_{2}|)<\varepsilon <{%
\frac{1}{2}}(u_{1}+u_{2}+|\mathcal{A}_{1}-\mathcal{A}_{2}|). 
\]%
When the inequality (\ref{eq19}) is not satisfied (figure \ref{Fig4}b
corresponds to this case), any horizontal line $\varepsilon =const$ crosses
some green region, i.e., there are always waves that can penetrate through
the barrier.

What is important for potential applications is that the parameters of the
barrier can be easily controlled by the applied voltage. Depending on the
voltage, \textit{this barrier can be either semitransparent} (Fig.~\ref{Fig4}b) \textit{or opaque} [with bound states made of surface waves (Fig.~\ref{Fig4}a)].

\section{Double barriers can produce waveguides}

Under certain conditions, two barriers separated by a distance $d$ form a 
\textit{waveguide}. Two barriers divide the infinite graphene sheet into
tree domains: left and right infinite half-planes, and the strip (of width $%
d $) bounded by the barriers. The potentials (electric and magnetic) are
constant within each domain. The values of the potentials are denoted by the
sub-indices $\ell ,$ $r$, and $c$, respectively, for left, right, and
middle. Each domain is represented by its own division of the $(\kappa
_{\perp },\,\varepsilon )$-plane in sectors with propagating and
non-propagating (in the given domain) waves, as it was described in the
previous section. The situation is similar to those presented in Fig.~\ref%
{Fig4}a,b, with only one difference: now there are tree Dirac points ($-%
\mathcal{A}_{\alpha },\,u_{\alpha }$) ($\alpha =\ell ,c,r$) with
``cones'' that divide the plane on several
sectors. Hereafter, the yellow-colored sectors correspond to the central ($c$%
) graphene domain, the blue-colored and violet-colored sectors correspond to
the right ($r$) and left ($\ell $) graphene domains, respectively. All
overlaps of the sectors will be marked by dark blue ($\ell $--$r$ overlap),
light green for the $r$--$c$ overlap, green for $\ell $--$c$ overlap, and
dark green for $r$--$c$--$\ell $ overlap.

The waveguide eigenmodes are non-propagating, evanescent waves in the left
and right infinite graphene half-planes. This means that the corresponding
points in the $(\kappa _{\perp },\,\varepsilon )$-plane are located 
outside of the colored sectors which are generated by the $r$ and $\ell $
Dirac points, i.e., in either yellow or white regions of the $(\kappa
_{\perp },\,\varepsilon )$-plane.

To simplify the presentation, we will mainly consider symmetric cases, $|u_{\ell }|=|u_{r}|,$ and $|\mathcal{A}_{\ell }|=|\mathcal{A}_{r}|$ and
assume, without loss of generality, that $u_{c}=\mathcal{A}_{c}=0$.

\subsection{Equal scalar and vector potentials: $u_{\ell }=u_{r}$ and $\mathcal{A}_{\ell }=\mathcal{A}_{r}$}

We first consider the case when $u_{\ell }=u_{r}$ and $A_{\ell }=A_{r}$. Setting $u_{\ell }=u_{r}=u_{2}$ and $\mathcal{A}_{\ell }=\mathcal{A}_{r}=\mathcal{A}_{2}$ (the blue-colored sectors
correspond to the $r$ and $\ell $ equivalent graphene domains), and setting $u_{c}=u_{1}=0$ and $\mathcal{A}_{c}=\mathcal{A}_{1}=0$ for the central
domain, one can now use Fig.~\ref{Fig4}a,b for describing the double-barrier
structure.

\begin{figure}[tbh]
\centering \scalebox{0.45}{\includegraphics{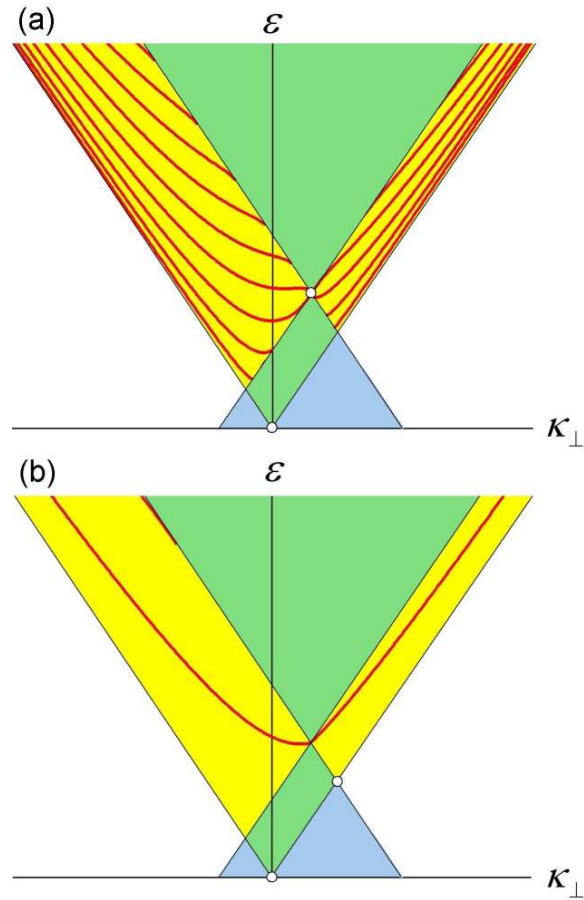}}
\caption{(color online) Waveguide spectrum $\varepsilon(\kappa_\perp)$ (red lines) when the inequality (\ref{eq19}) is not valid.
(a) for large spacing $d$ between the barriers; (b) for small spacing $d$
between the barriers.}
\label{Fig5}
\end{figure}

When the inequality (\ref{eq19}) is not valid (as in Fig.~\ref{Fig4}b),
there are waveguide modes which are similar to the modes in the usual
dielectric waveguides. The spectra of these modes are shown by the red lines
in Fig.~\ref{Fig5}a,b for different values of the distance $d$ between the
barriers (waveguide walls).

As in a dielectric waveguide, the total internal reflection (TIR) phenomenon
is responsible on the wave confinement in the central graphene domain.
However, in contrast to usual dielectric waveguides, in a certain energy
range there are two separate regions of $\kappa _{\perp }$ where TIR occurs.
Decreasing the distance $d$ between the barriers shifts the spectrum to
higher-energies, with the exception of the lowest mode, which crosses the
point ($-\mathcal{A}_{\ell },\,u_{\ell }$) for whatever small values of $d$
(Fig.~\ref{Fig5}b).

When the inequality (\ref{eq19}) is valid, there are confined waveguide
modes in the yellow regions in Fig.~\ref{Fig4}a. In addition to these
``ordinary'' modes, there are two
``extraordinary'' modes in the white region
between two Dirac points where the surface wave exists. In contrast to the
ordinary modes that appear due to the TIR, the extraordinary modes are
formed by two coupled surface waves propagating along the barriers.

\begin{figure}[tbh]
\centering \scalebox{0.45}{\includegraphics{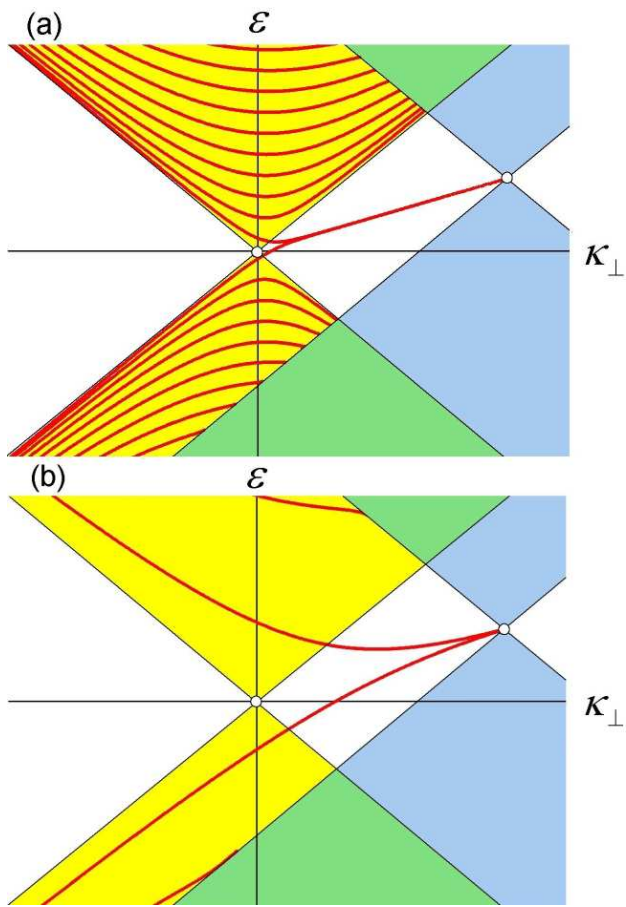}}
\caption{(color online) Waveguide spectra (red lines) when the inequality (%
\ref{eq19}) is valid. (a) large spacing $d$ between
the barriers. The isolated red line in (a) involves two extraordinary modes,
corresponding to edge states, which merge in the isolated red line because
these two curves are very close to each other. The difference between these
nearly-overlapping modes is visible only when the modes are transformed into
ordinary ones. (b) small spacing $d$ between the barriers. Note
that the spacing between the energy levels increases when the spacing
decreases.}
\label{Fig6}
\end{figure}

In Fig.~\ref{Fig6}a, the blue 2D ``cone''
on the right represents the overlap of two 2D ``cones'' corresponding to the left and right graphene
semiplanes. The isolated red solid line originated at the right Dirac point
actually represents two dispersion curves, which in this instance are
indiscernible. When these two nearly-overlapping curves approach the left
Dirac point, they separate moving in opposite directions: one towards
positive $\varepsilon $, the other towards negative $\varepsilon $.
In other words, the two coupled surface waves propagating along the barriers
(red solid curve emanating from the right Dirac point) transform smoothly
into the standard ordinary modes, when approaching the left Dirac point.
Examples of waveguide spectra are shown in Fig.~\ref{Fig6}a,b for different
values of the distance $d$.

The particular case $u_{\ell }=u_{r}=0$ has been considered in \cite{Pereira, Masir1, Masir3, SGhosh}, and the case $A_{\ell }=A_{r}=0$ has been
considered in \cite{Yampolskii1,Yampolskii2}.

The eigenspectrum of both ordinary and extraordinary modes is defined by the
requirement that a round trip phase 
\[
\phi =2\mathrm{Re}(\kappa _{\parallel })d+\phi _{\ell }+\phi _{r}, 
\]%
($\phi _{\ell }$ and $\phi _{r}$ are the phase shifts at the reflection from
the left and right barriers) is either equal to zero (extraordinary
evanescent modes) or to a multiple of $2\pi n$, $n=1,2,\ldots$ (ordinary
propagating modes).

There is an important difference between the ordinary and extraordinary
modes: when the distance between the barriers is large enough, the
extraordinary modes are \textit{unidirectional} waves (see Fig.~\ref{Fig6}%
a). This means that all the waves (in the extraordinary mode existence
region) propagate in one direction, making this mode \textit{resistant to
backscattering}, and therefor robust against $y$-dependent
disorder. Indeed, any rather smooth obstacle in the waveguide cannot
produce counter-propagating waves. 
As to the fluctuations of the height of the
potentials (in $x$-direction), in principle, they can
affect dramatically the propagation \textit{across} the barriers, but have
practically no influence on the waveguide eigenmodes. Note that the properties of the waveguide
eigenmodes are easily controlled by the electrostatic potentials.

\begin{figure}[tbh]
\centering \scalebox{0.45}{\includegraphics{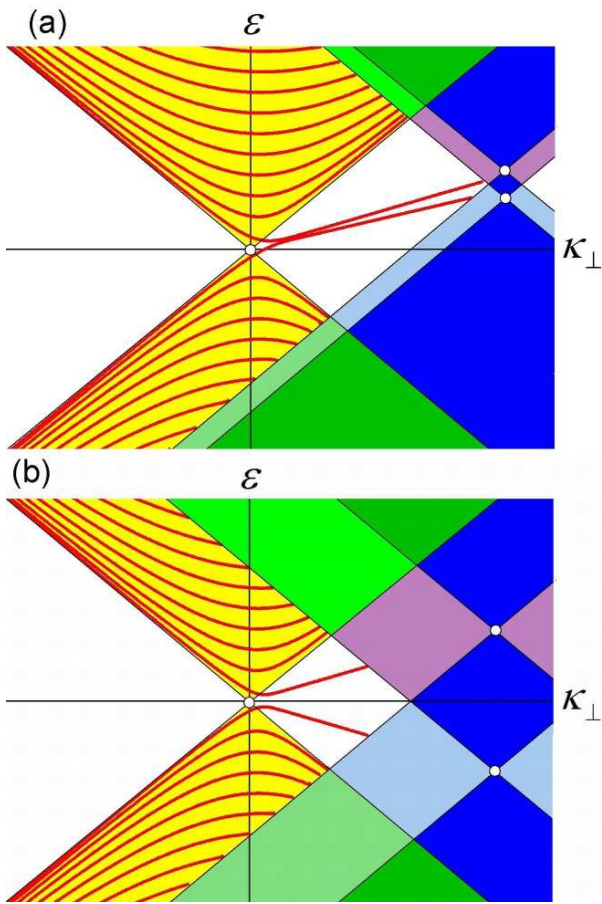}}
\caption{(color online) Waveguide spectra for the cases discussed in the
subsection IV A. (a) for $u_\ell\neq u_r$, $\mathrm{sgn}(u_\ell)=\mathrm{sgn}%
(u_r)$; (b) for $u_\ell=- u_r$. The difference between the two extraordinary
modes is distinctly visible. For the case $\mathrm{sgn}(u_\ell)=-\mathrm{sgn}%
(u_r)$, the modes in (b) have opposite-directed group velocities.}
\label{Fig6a}
\end{figure}

The closeness of two extraordinary waves spectra (central isolated red line
in Fig.~\ref{Fig6}a) stems from the chosen symmetry of the waveguide
potentials $u_{\ell }=u_{r}$. When this symmetry is broken, $u_{\ell }\neq
u_{r}$, the difference between the spectra is clearly visible (two central
red lines in Fig~\ref{Fig6a}a). When the potentials $u_{\ell }$ and $u_{r}$
have opposite signs, surface waves propagate along the barriers in opposite
directions. As a result, two extraordinary modes not only have separated
spectra, but propagate in opposite directions, as it is shown in Fig~\ref{Fig6a}b. The backscattering is also suppressed in this case because the
counter-propagating waves are localized mainly near the corresponding
barriers.

\subsection{Equal scalar potentials and antiparallel vector potentials: $%
u_{\ell }=u_{r}$ and $\mathcal{A}_{\ell }=-\mathcal{A}_{r}$}

In this subsection, we will concentrate on the case when Eq.~(\ref{eq19}) is
valid. The instance when the inverse of the (\ref{eq19}) inequality is valid
is similar to one considered in subsection IV\ A, just with a more complex
division of the ($\kappa _{\perp },\,\varepsilon $)-plane in regions.

\begin{figure}[tbh]
\centering \scalebox{0.45}{\includegraphics{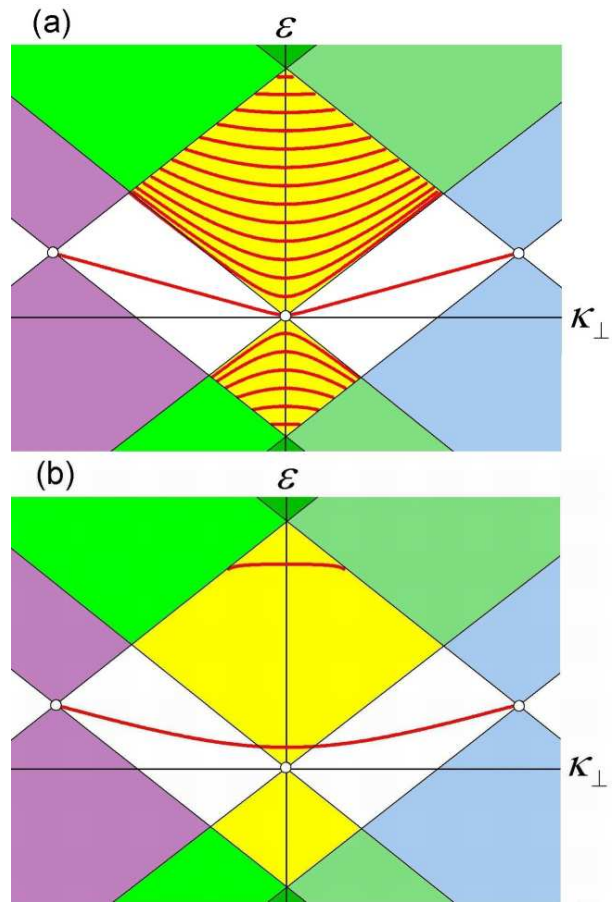}}
\caption{(color online) Waveguide spectra when the scalar $u_\ell=u_r$ and
vector $\mathcal{A}_\ell=-\mathcal{A}_r$ potentials satisfy these relations.
(a) large distance $d$ between the barriers; and (b) small distance between
the barriers. When the distance $d$ decreases, the ordinary modes are pushed
out from the yellow regions and only the extraordinary mode remains in a broad energy gap.}
\label{Fig7}
\end{figure}

The condition $u_{\ell }/\mathcal{A}_{\ell }=-u_{r}/\mathcal{A}_{r}$ means
that the surface waves along the left and right barriers propagate in
opposite directions making the waveguide eigenmodes spectra symmetric, as
shown in Fig.~\ref{Fig7}a,b.

In contrast to the case of equal potentials (subsection IV.b) the spectra of
ordinary and extraordinary modes are separated: there is no transformation
of one kind of mode into another, as it is seen in Fig.~\ref{Fig6} and Fig.~%
\ref{Fig6a}. Note that the coupling between two surface waves, which
produces the extraordinary mode does not split the mode energy, as it
happens when $u_{\ell }=u_{r}$ and $\mathcal{A}_{\ell }=\mathcal{A}_{r}$.

\begin{figure}[tbh]
\centering \scalebox{0.8}{\includegraphics{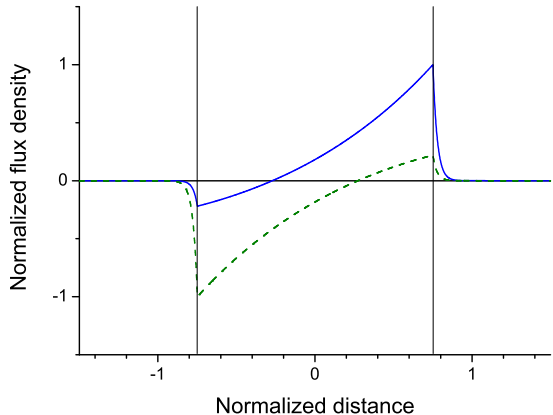}}
\caption{(color online) Normalized current density of the extraordinary
modes (counter-propagating surface waves). Solid line: $\kappa_\perp>0 $; dashed line: $\kappa_\perp<0$. The vertical lines
indicate the positions of the barriers. The counter-propagating currents are
indeed spatially separated.}
\label{Fig8}
\end{figure}

There is an energy range where only extraordinary modes form the charge flux
along the waveguide. Despite the fact that the extraordinary modes are not
unidirectional, the backscattering is also suppressed in this case. The
reason for this is the spatial separation of regions with opposite direction
of the flux: the wave functions that correspond to opposite directions of
flux are localized near different (left or right) barriers. This is
illustrated in Fig.~\ref{Fig8}, where the normalized flux densities for
extraordinary eigenmodes with opposite wave vectors $\kappa _{\perp
1}=-\kappa _{\perp 2}$ (opposite total fluxes) are shown.

Decrease of the spacing, $d\rightarrow 0$, broadens the energy gap where
only extraordinary mode exists (see Fig.~\ref{Fig8}b). The dispersion curve
is then flattened and tends to the line $\varepsilon (\kappa _{\perp
})=u_{\ell }=u_{r}$.

\subsection{Scalar potentials of opposite signs and antiparallel vector
potentials: $u_{\ell }=-u_{r}$ and $\mathcal{A}_{\ell }=-\mathcal{A}_{r}$}

\begin{figure}[tbh]
\centering \scalebox{0.45}{\includegraphics{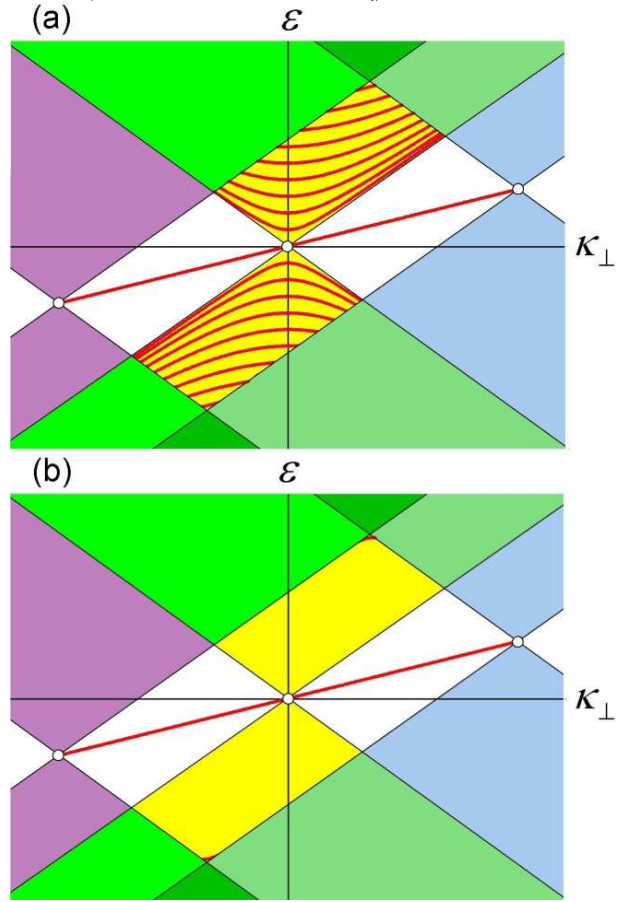}}
\caption{(color online) Waveguide spectra for the case described in
subsection IV.C, with opposite scalar and vector potentials. (a) large
spacing between the barriers; (b) small spacing between the barriers.}
\label{Fig10}
\end{figure}

When $u_{\ell }=-u_{r}$, $\mathcal{A}_{\ell }=-\mathcal{A}_{r}$ and the
inequality (\ref{eq19}) holds, both barriers support surface waves with the
same direction of the charge flux. Accordingly, the flux which is associated
with the extraordinary waves is unidirectional, i.e., is independent on sign
of $\kappa _{\perp }$. Examples of waveguide spectra are shown in Fig.~\ref{Fig10}a,b. When the distance $d$ is small enough, only extraordinary waves
are confined by the waveguide.

Comparing Figs.~\ref{Fig10}a and \ref{Fig10}b, one can see that, unlike the
ordinary modes, the spectrum of the extraordinary modes is \textit{independent} \textit{of the distance} $d$ \textit{between barriers}. There
is a cut-off energy $\varepsilon _{0}$ for ordinary modes, i.e., the minimal
modulus $|\varepsilon |$ of the energy, for which the condition $\phi =2\pi $
is satisfied (the cut-off energy is the analog of the cut-off frequency of
conventional waveguides). The cut-off energy increases when the distance $d$
decreases, pushing out the ordinary modes from their existence region
(yellow regions in Fig.~\ref{Fig10}a,b). It is interesting to note that in
the graphene waveguide with variable width $d(y)$, the extraordinary mode
can penetrate through an arbitrary narrow part of the waveguide, whereas all
ordinary modes are reflected from it. A similar effect (penetration of the
electromagnetic wave through the waveguide waist) is typical to waveguides
filled with a metamaterial with a near-zero dielectric permittivity \cite{Engheta}. A rather narrow waveguide is a single-mode waveguide, although
its transport properties differ strongly from the usual single-mode
waveguides. The transmission through this waveguide (in the frame of the
model we use) is always perfect, irrespectively of smooth irregularities
present on the waveguide. As in the single barrier case, the reason is the
absence of counter-propagating waves that makes reflection impossible.

\section{Conclusions}

We have shown that crossed magnetic and electric fields applied to a narrow
strip (electromagnetic barrier) on a graphene sheet can form \textit{a
unidirectional conducting channel (quantum wire)} whose properties are
easily \textit{tunable} by the voltage applied across the strip. The
eigenmode of this channel is characterized by a linear dispersion and
represents a one-way propagating wave. This unique property \textit{prohibits backscattering} and therefore makes the mode \textit{resistant to
the scattering by impurities}. The classical analogy of this mode is the 
\textit{drift }of a charged particle in crossed electric and magnetic
fields. The transverse-localized mode exists and propagates longitudinally
along the barrier with the drift velocity $v_{y}=v_{d}\equiv cE_{x}/H_{z}$
if and only if this velocity $v_{d}$ is smaller than the Fermi velocity $v_{F}$: i.e., $v_{d}< v_{F}$. While one barrier forms a wire, two such
barriers produce a waveguide whose eigenfunctions consist of set of ordinary
and extraordinary waves. The ordinary waves are characterized by the
quantization of their transverse wave numbers, while the extraordinary waves
are formed by two coupled surface waves propagating along the waveguide
walls (barriers). The rather narrow waveguide has only extraordinary
eigenmode. Depending on the parameters of the walls this extraordinary
eigenmode can be either uni- or bi-directional. In the bi-directional case
the regions with opposite directions of current flows are spatially
separated, preventing backscattering and making even the bi-directional mode
resistant against the scattering by impurities.

\section*{Acknowledgments}

FN gratefully acknowledges partial support from the National Security Agency
(NSA), Laboratory of Physical Sciences (LPS), Army Research Office (ARO),
and National Science Foundation (NSF) under Grants No. 0726909 and No.
JSPS-RFBR 06-02-91200.


\begin{thebibliography}{99}
\bibitem{Castro1} A.H. Castro Neto, F. Guinea, N.M.R. Peres, K.S. Novoselov,
and A.K. Geim, Rev. Mod. Phys. \textbf{81}, 109 (2009).

\bibitem{Geim1} A.K. Geim and K.S. Novoselov, Nature Materials \textbf{6},
183 (2007).

\bibitem{Novoselov2} K. Novoselov et al., Nature \textbf{438}, 197 (2005).


\bibitem{Katsnelson} M.I. Katsnelson, Materials Today \textbf{10}, 20 (2007).

\bibitem{Trauzettel} B. Trauzettel, D.V. Bulaev, D. Loss, and G. Burkard,
Nature Physics \textbf{3}, 192 (2007).

\bibitem{Berger} C. Berger, Z. Song, T. Li et al. J. Phys. Chem. B, \textbf{%
108}, 19912 (2004).

\bibitem{Son} Y.-W. Son, M.L. Cohen, and S.G. Louie, Nature \textbf{444},
347 (2006).

\bibitem{Klein} O. Klein, Z. Phys. \textbf{53}, 157 (1929).

\bibitem{Katsnelson2} M.I. Katsnelson, K.S. Novoselov, and A.K. Geim, Nature
Physics \textbf{2}, 620 (2006).

\bibitem{DeMartino1} A. De Martino, L. Dell\'{}Anna, and R. Egger, Phys. Rev.
Lett. \textbf{98}, 066802 (2007).

\bibitem{DeMartino2} A. De Martino, L. Dell\'{}Anna, and R.
Egger, Solid Stat. Commun. \textbf{144}, 547 (2007).

\bibitem{Pereira} V.M. Pereira and A.H. Castro Neto, Phys. Rev. Lett. 
\textbf{103}, 046801 (2009).

\bibitem{Guinea} F. Guinea, M.I. Katsnelson, and A.K. Geim, Nature Physics \textbf{6}, 30 (2010); arXiv:cond-mat/0909.1787

\bibitem{Guinea-2}  M.M. Fogler, F. Guinea, and  M.I. Katsnelson,
Phys. Rev. Lett. \textbf{101}, 226804 (2008).

\bibitem{Masir2} M. Ramezani Masir, P. Vasilopoulos, A. Matulis, and F. M.
Peeters, Phys. Rev. B \textbf{77}, 235443 (2008).

\bibitem{Masir1} M. Ramezani Masir, A. Matulis, and F.M. Peeters, Phys. Rev.
B \textbf{79}, 155451 (2009).

\bibitem{Kormanyos} A. Korm\'{a}nyos, P. Rakyta, L. Oroszl\'{a}ny, and J.
Cserti, Phys. Rev. B \textbf{78}, 045430 (2008).

\bibitem{Park} S. Park and H.-S. Sim, Phys. Rev. B \textbf{77}, 075433
(2008).

\bibitem{Muller} J.E. M$\ddot{\mathrm{u}}$ller, Phys. Rev. Lett. \textbf{68}%
, 385 (1992).

\bibitem{Orozlany} L. Oroszl\'any, P. Rakyta, A. Korm\'anyos, C.J. Lambert,
and J. Cserti, Phys. Rev. B \textbf{77}, 081403(R) (2008).

\bibitem{Ghosh1} T.K. Ghosh, A. De Martino, W. H$\ddot{\mathrm{a}}$usler, L.
Dell\'{}Anna, and R. Egger, Phys. Rev. B \textbf{77}, 081404(R) (2008).

\bibitem{Peeters1} F.M. Peeters and A. Matulis, Phys. Rev. B \textbf{48},
15166 (1993).

\bibitem{Peeters2} J. Reijniers and F.M. Peeters, Phys. Rev. B. \textbf{63},
165317 (2001).

\bibitem{Peeters3} J. M. Pereira Jr., F.M. Peeters, and P. Vasilopoulos,
Phys. Rev. B \textbf{75}, 125433 (2007).

\bibitem{Titov} M. Titov, Eur. Phys. Lett. \textbf{79}, 17004 (2007).

\bibitem{BliokhPRB} Y.P. Bliokh, V. Freilikher, S. Savel'ev, and F. Nori,
Phys. Rev. B \textbf{79}, 075123 (2009).

\bibitem{Rozhkov} A.V. Rozhkov, S.S. Savel'ev, and F. Nori, Phys. Rev. B 
\textbf{79}, 125420, (2009).

\bibitem{Feng} F. Zhai and K. Chang, Phys. Rev. B \textbf{77}, 113409 (2008).

\bibitem{Kubrak}  R. Kubrak, A. Neumann, B.L. Gallagher, P.C. Main,
M. Henini, C.H. Marrows, and B.J. Hickey, J. Appl. Phys. \textbf{87}, 5986 (2000).

\bibitem{Crerchez}  M. Cerchez, S. Hugger, T. Heinzel, and N.
Schulz, Phys. Rev. B \textbf{75}, 035341 (2007).

\bibitem{Bader}  S.D. Bader, Rev. Mod. Phys. \textbf{78}, 1 (2006).

\bibitem{Masir3} M. Ramezani Masir, P. Vasilopoulos, and F. M. Peeters, New
J. Phys. \textbf{11}, 095009 (2009).

\bibitem{SGhosh} S. Ghosh and M. Sharma, J. Phys.: Condens. Matter \textbf{21}, 292204 (2009); S. Ghosh and M. Sharma, arXiv:cond-mat/0806.2951 (2009).

\bibitem{Yampolskii1} V.A. Yampol'skii, S. Savel'ev, and F. Nori, New J.
Phys. \textbf{10}, 053024 (2007).

\bibitem{Yampolskii2} V.A. Yampol'skii, S.S. Apostolov, Z.A.
Maizelis, A. Levchenko, and F. Nori, arXiv:cond-mat/0903.0078 (2009).

\bibitem{Engheta} M. Silveirinha and N. Engheta, Phys. Rev. Lett. \textbf{97}, 157403 (2006).


\end{thebibliography}
\end{document}